\documentstyle[12pt]{article}
\begin{document}
\begin{titlepage}
\begin{flushright} UTPT-93-08 \\ hep-ph/9304309 \\ September 1993
\end{flushright}
\vspace{24pt}
\begin{center} {\LARGE Linear and higher order power corrections\\  in
semileptonic $B$ decays}\\ \vspace{40pt} {\large B.
Holdom\footnote{holdom@utcc.utoronto.ca} and M.
Sutherland\footnote{marks@medb.physics.utoronto.ca}}
\vspace{0.5cm}

{\it Department of Physics\\ University of Toronto\\
Toronto, Ontario\\Canada M5S 1A7}

\vspace{12pt}

\begin{abstract}
In previous work we have developed a relativistic quark model of mesons
which is consistent with all QCD constraints at zeroth and first order
in the heavy-quark expansion.   Here we obtain first-order model
predictions for the differential decay  spectrum, the forward-backward
asymmetry $A_{FB}$ and the $D^{\ast}$ polarization parameter $\alpha$
in the decay $B\rightarrow  D^{\ast}\ell\nu$.  We compare these with
the predictions of QCD sum rules at first order.  The model suggests
why the corrections allowed at first order are small, concurrently with
substantial corrections at second order.
\end{abstract}
\end{center}  \end{titlepage} \pagebreak \setcounter{footnote}{0}
{\bf \noindent I. INTRODUCTION} \vspace{2ex}

We have recently developed a relativistic model for $B$-meson
semileptonic decays
$B\rightarrow D^{(\ast)}\ell\nu$, in which the  hadronic matrix
elements are represented by quark loop graphs with  damping factors at
the $Qq$-meson vertices
\cite{HS1,HS2}.  These  have the form \begin{equation}
F_{P,V}(k)=\frac{Z_{P,V}^{2}}{-k^{2}+\Lambda_{P,V}^{2}} \label{df}
\end{equation} where $k$ is the momentum of the light quark and $P$ and
$V$  denote pseudoscalar and vector mesons.  These vertices together
with standard quark propagators determine what we will call the  ``full
model", in which no reference is made to any expansion in  inverse
powers of heavy-quark masses.  The only parameters of the  full model
are the heavy- and light-quark masses, in terms of which  the constants
$\Lambda_{P,V}$ and $Z_{P,V}$ are fully determined  by requiring that
the meson self-energy functions vanish and have  unit slope at the
physically-measured meson masses.

We may expand all quantities in the model in inverse powers of  heavy
quark masses.  Severe constraints on the form of the expansion follow
from  QCD using the heavy-quark effective theory (HQET) \cite{Geo}.  In
the  heavy-quark limit, four of the six form factors for
$B\rightarrow D^{(\ast)}\ell\nu$ become equal to a single universal
Isgur-Wise  function, while the others vanish \cite{IW}.  The twelve
{\it a priori}  independent $1/m_{c}$ and $1/m_{b}$ corrections to  the
form factors are given  by specific linear combinations of the
Isgur-Wise function and four additional universal functions
\cite{Luke,NeuRieck}.  The actual  shapes of these universal functions
are model-dependent, but some have model-independent  values at zero
recoil.  We have shown in  \cite{HS1} that all such constraints are
satisfied by our model at first order.

The HQET restrictions arising at second order ($1/m_{Q}^{2}$) are
available in \cite{FalkNeu1}.  We have not yet demonstrated
consistency with these results, as our model for mesons is rather
unwieldy at this order.  On the other hand we  have developed a similar
model for $\Lambda_{Q}$ baryons  \cite{HS5}.  It is more tractable than
the meson case and is fully  consistent with the HQET constraints
presented in \cite{FalkNeu2} up  to and including second order.  Our
approach for mesons follows the  same principles and we thus conjecture
that it is also consistent at  second order.

As was shown in \cite{HS2}, the main feature of the full model  before
expansion is the presence of rather large deviations from the heavy
quark limit predictions for the zero-recoil values of the form factors
$h_{+}$ and  $h_{A_{1}}$, defined in eqs. (\ref{formfacs1}) and
(\ref{formfacs2}) below.  These quantities play a crucial role in the
determination of the Kobayashi-Maskawa element $V_{cb}$, as discussed
in \cite{Neu1}.   Both are equal to 1 in the  heavy-quark limit and
neither receives corrections at order $1/m_{Q}$ (Luke's theorem
\cite{Luke}).  The corrections we found were traced to the effects of
hyperfine mass splitting.   For the choice of quark masses in
\cite{HS2} we found values of 10 to 15  per cent for the corrections at
all orders, with the corrections occurring purely at second order
making up about $2/3$ of these full corrections.  This translates
almost directly into a 10 to 15 per cent model dependence in $V_{cb}$.

We believe that it is currently of interest to obtain estimates of the
corrections beyond first order in $1/m_Q$ in any model, such as ours,
which is consistent with the heavy-quark symmetries.  Many of the
previously popular models \cite{ISGW,BSW,KS2} cannot be used since they
do not obey the  symmetry  constraints, as described in
\cite{Neurevue}.  Of course the question remains as to how well our
model resembles QCD.  Perhaps the main question has to do with
confinement; our representation of the  effects of confinement is to
simply ignore the imaginary parts arising from our free quark loop
calculations.  We emphasize again, though, that this procedure is
consistent with Ward identities and heavy-quark symmetries \cite{HS1}.

Another approach \cite{FalkNeu1} has been to carefully study the
structure of the second-order corrections in the HQET.  Here the
question boils down to the estimation of matrix elements of various
operators.  Sum rules  have been applied to the first-order
corrections, but the estimates of the  second-order corrections are
much less sophisticated.  In \cite{FalkNeu1}  some matrix elements are
estimated using the Isgur-Scora-Grinstein-Wise (ISGW) nonrelativistic
quark  model
\cite{ISGW}, but this model does not include hyperfine splitting
effects.  In fact when hyperfine splitting effects are turned off in
our model (i.e. when
$g=h=0$ in (\ref{aaa}) below) we find order $1/{m}_{Q}^{2}$
contributions to ${h}_{+}$ and ${h}_{{A}_{1}}$ at zero recoil which are
very similar to those of the ISGW model.

Other matrix elements involve double insertions of the  chromomagnetic
moment operator, and these are neglected in \cite{FalkNeu1} since the
single insertions appearing at first order are observed to be small.
But these double insertion matrix elements are precisely the ones which
are large in  our model \cite{HS2}, and they are large concurrently with
small values for the  single insertion matrix elements.  We will argue
below that the small size of the first-order corrections to semileptonic
decay amplitudes is the exception rather than the rule.  The suggested
mechanism that suppresses the first-order corrections does not apply to
higher order corrections.  Even at first order large corrections are
possible; for example our model and sum rules agree that the heavy meson
decay constants receive much larger first-order corrections than the
semileptonic decay amplitudes \cite{HS1}.

Until more reliable estimates of the second-order corrections appear,
our model gives some indication of the possible theoretical uncertainty
in  the extraction of $V_{cb}$.  One of the objects of this paper is to
provide  a set of model predictions to be compared to future data.  In
particular we will define a ``first-order model" in which only the
first order corrections are kept, and we consider quantities which are
fairly insensitive to the higher order corrections.  We compare these
results with QCD sum rules
\cite{NeuQCD} and find that the differences may be large enough to make
the two first-order models experimentally distinguishable.  We will
also see from present data that drastic modifications to the model
cannot be tolerated.

In the last part of the paper we shall develop more understanding of
the higher order corrections found in the full model.  We will stress
the consequences of holding the physical meson masses fixed in the
full model, in particular constraints on $m_c$ and $m_b$.

\vspace{2ex}

{\bf \noindent II. THE FIRST-ORDER MODEL} \vspace{2ex}

The first-order model has three parameters, besides the quark masses.
The heavy-quark limit $m_{Q}\rightarrow \infty$  is characterized
uniquely by the ratio $\Lambda/m_{q}$, where  $m_{q}$ is the light
quark mass and $\Lambda$ is the common value  of $\Lambda_{P,V}$ in the
heavy-quark limit.  The other two  parameters $g$ and $h$ characterize
the approach to the heavy  quark limit via \begin{equation} {\Lambda
}_{P,V}=\Lambda \left({1-(\delta_{P,V} h+ g){\frac{\Lambda
}{{m}_{Q}}}}\right)\label{aaa} \end{equation} where ${\delta}_{P}=3$ and
${\delta}_{V}=-1$.\footnote{The definition of $g$ and $h$ here differs
from that in \cite{HS1}.}  The bulk of hyperfine mass splitting is
contained in  $h$, a positive value of which drives the pseudoscalar
mass down and the vector mass up by amounts in the ratio of 3 to 1.
But not all effects of the heavy-quark spin-symmetry breaking are
described by $h$; some breaking is intrinsic to the relativistic quark
loop and it is present when $h=0$.  The three parameters of the first
order model are free parameters, and the consistency with heavy-quark
symmetry is valid for any values of these parameters.

For the purpose of computing physical predictions with the first
order model, numerical values of its parameters may be estimated as
follows.  As in \cite{HS1,HS2}, we choose $m_{b}=4.8$ GeV,
$m_{c}=1.44$ GeV and $m_{q}=250$ MeV.   The first-order model  then
yields  expressions for the $B$, $B^{\ast}$, $D$ and $D^{\ast}$ masses
in  terms of $\Lambda$, $g$ and $h$.  These we adjust to fit the  four
masses, yielding an optimal mass spectrum when  $\Lambda=667$ MeV,
$g=-0.13$, and $h=0.19$.  In this way the model is fixed once the
quark masses are given, and we find that changes in the first-order
results are minor for any reasonable choice of quark masses.

The mass difference between the meson and the heavy quark in the  heavy-
quark limit is denoted by $\overline{\Lambda}$  \cite{FNL}; it is
directly related to $\Lambda$ \cite{HS1}. With the above  values of
$\Lambda$ and  $m_{q}$, we have  $\overline{\Lambda}=504$ MeV, which
coincides with the QCD sum rule estimate in \cite{NeuQCD}.  In
addition, the Isgur-Wise function from  this model is numerically very
similar to one given in a sum rule analysis \cite{NeuSR}:
\begin{equation} \xi(\omega)^{\rm s.r.}=
\left(\frac{2}{\omega+1}\right)^{1.88+ \frac{0.69}{\omega}}.
\end{equation} The slope at $\omega=1$ is $-1.28$.\footnote{More recent
sum rule analyses favor a less negative slope \cite{Neurevue}.}

As mentioned above, four additional universal functions, $\chi_{1}$,
$\chi_{2}$, $\chi_{3}$, and $\xi_{3}$, appear at first order
\cite{Luke,FNL}.  Their values are plotted for the above parameter
values in Fig. 1.   All are relatively small compared with the
Isgur-Wise function,  which is 1 at zero recoil and approximately 0.6
at $\omega=1.5$.  In fact we can begin to see how it is that
dimensionless quantities appearing at first order are small compared to
unity.  $\chi_{2}$ and $\xi_{3}$ are spin-symmetry violating and
conserving, respectively, but they are both independent of the
parameters $g$ and $h$ for any $\omega$ \cite{HS1}.  They are therefore
independent of the ``wave  function" distortions described by $g$ and
$h$ which are required to fit the physical meson masses.  $\chi_{1}$ is
spin-symmetry conserving and therefore depends only on $g$ while
$\chi_{3}$ is spin-symmetry violating and depends only on $h$.  But
both $\chi_{1}$ and $\chi_{3}$ are constrained to be zero at zero
recoil by Luke's Theorem, and thus for the physical range of $\omega$
they remain small.  There is thus no quantity at first order which is
both sensitive to ``wave function" distortions and nonvanishing at zero
recoil.  In contrast, there are such dimensionless quantities appearing
at second order in the heavy-quark expansion \cite{HS2}.  Such
quantities can be expected to be of order unity, and they thereby give
corrections which appear to be large when compared to the first-order
corrections.

Since $\chi_{2}$ and $\xi_{3}$ are independent of the parameters $g$
and $h$ the following combinations of  $B\rightarrow D^{\ast}$ form
factors are especially interesting: \begin{equation}
R_{1}=\frac{h_{V}}{h_{A_{1}}}\approx  1+ \frac{1}{\omega+1}\left(
\frac{\overline{\Lambda}}{m_{c}} +  \frac{\overline{\Lambda}}{m_{b}}
[1-2\frac{\xi_{3}}{\xi}]\right) \label{approx1} \end{equation}
\begin{eqnarray} R_{2}&=&\frac{
h_{A_{3}}+(m_{D^{\ast}}/m_{B})h_{A_{2}}}{h_{A_{1}}} \nonumber \\
&\approx& 1 + \frac{1}{\xi}\left\{ \frac{\overline{\Lambda}}{m_{c}}
\left( \frac{-\xi_{3}}{1+\omega}-2\chi_{2} \right)
+\frac{\overline{\Lambda}}{m_{b}} \left(
\frac{-3\xi_{3}}{1+\omega}+2\chi_{2} \right) \right\}. \label{approx2}
\end{eqnarray}  These quantities have also been stressed elsewhere
\cite{NeuQCD}.  Both $R_{1}$ and $R_{2}$ are nearly constant across the
spectrum.   The main difference with QCD sum rules is that our $\xi_{3}$
is much smaller.  Our result $R_{1}\approx 1.3$ is still in agreement
with QCD sum rules due to the $1/m_b$ suppression of the second term.
But $R_{2}\approx  1.0$ differs somewhat from the value of 0.8 found in
QCD sum rules  \cite{Neurevue}.

Indeed we find, in contrast to the Isgur-Wise function, that all four
universal functions arising at first order in our model are rather
different from those in QCD sum rules, and that they lead to different
physical predictions.  For example, we may consider the differential
$B\rightarrow D^{\ast}$ spectrum in $\omega$.  In the heavy-quark
limit this goes over to $|V_{cb}|^{2}g(\omega)\xi(\omega)^{2}$ where
$g(\omega)$ is a known function of the meson masses and includes
short distance QCD corrections.  (We include short-distance QCD
corrections \cite{NeuQCD} in all subsequent computations.)  Dividing
the spectrum by $g(\omega)$ and taking the square root then yields
$|V_{cb}|f(\omega)$ where $f(\omega)$ goes over to $\xi(\omega)$  in
the heavy-quark limit.   The first-order model predicts  $f(\omega)$,
and we obtain $|V_{cb}|$ by fitting to the data.

This raises an important point.  The normalization $f(1)=1$ at first
order is model-independent, so it would be possible to obtain
$|V_{cb}|$ is a model-independent way if there were data at zero
recoil.  But the differential spectrum  vanishes at zero recoil, so we
require an extrapolation.  This extrapolation can only be accomplished
by fitting  some functional form to the data; this functional form is
model-dependent  and hence so is $|V_{cb}|$.  This model-dependence
will diminish only as  the data improves.

We plot in Fig. 2 the first-order model results together
with  ARGUS \cite{ARGUS} and CLEO \cite{Borto} data for
$|V_{cb}|f(\omega)$.  We see that the shape of our curve is steeper
than that predicted by QCD sum rules, and that the present data data
favors the steeper curve.  At first order $f(1)=1$, so $|V_{cb}|$ may
simply be  read off as the intercept of the curve.  We  find
$|V_{cb}|=.042$ in our  model and $|V_{cb}|=.037$ in QCD sum  rules, a
difference of 13\%.   This  large discrepancy is due entirely to the
first-order corrections, since we  have used the same Isgur-Wise
function for both cases.

We next present results for two integrated quantities which directly
sample the data away from zero recoil: the $D^{\ast}$ polarization
parameter $\alpha$ and the forward-backward asymmetry   $A_{FB}$.
These are plotted versus the experimental lepton momentum cut in
Figs. 3 and 4.   The present experimental
values \cite{ARGUS,CLEO89,CLEO} of $A_{FB}$ and $\alpha$ are also
displayed.  In both cases, the predictions of our first-order model
differ from those of QCD sum rules; better data would make it
possible to distinguish between them.

\vspace{2ex}

{\bf \noindent III. THE FULL MODEL} \vspace{2ex}

If we consider the same quark masses as above then the full model
yields \cite{HS2} $h_{+}(1)=1.107$ and $h_{A_{1}}(1)=1.155$, in
contrast to their model-independent values of 1 at first order.  The
value of $|V_{cb}|$ is then no longer given by the intercept of the
full model curve shown in Fig. 2, and $|V_{cb}|$ in this
case is  $0.38$.  In contrast, the effects of these higher order
corrections manifest themselves only slightly in $\alpha$, as shown in
Fig.  3.  And as seen in Fig. 4 they cancel
out almost completely  in $A_{FB}$.

We now turn to a discussion of the sensitivity of the full model
results to the heavy-quark masses.  By fixing the  physical values of
the meson masses, the quantities $Z_{P,V}$  and   $\Lambda_{P,V}$ of
Eq. \ref{df} are fully determined for each meson, as  mentioned at  the
outset.  This points to a fundamental difference in  the
interpretation  of the full model versus the heavy-quark expansion.  In
the latter, the  meson masses vary as the  heavy-quark mass varies. But
in the  full model the physical meson masses  are held fixed.  Then
the  dependence of various quantities on the heavy-quark mass has no
connection with the standard dependence in the heavy-quark expansion.
In fact in the full model it will only be possible to produce physical
meson masses when the quark masses are within  some range.  This is a
feature of any realistic model of QCD for fixed hadron masses.   In
fact, the better the model, the more the quark masses should be
constrained to their true values.

We can capitalize on this fact and use the full model to constrain
$m_c$ and $m_b$.  An easily measurable quantity which has strong
dependence on these masses is the ratio of branching ratios
$B(B\rightarrow D)/B(B\rightarrow  D^{\ast})$.  We illustrate this by
plotting this ratio in Fig. 5 as a function of  $m_{c}$ at
fixed $m_{b}=4.8$ GeV (with no lepton momentum cut).   As the data
improves this mass dependence will translate into a constraint on $m_c$
as a function of $m_b$.

In fact, constraints on the heavy-quark masses already arise.  We
consider both $B\rightarrow  D^{\ast}$ and  $B\rightarrow D$ processes
and we find that the allowed region in the $m_{c}$-$m_{b}$ plane is
defined by \begin{eqnarray} m_{c}+\Lambda_{B}(m_{b};m_{q})>M_{D^{\ast}}
\label{Dstconstraint}  \\ m_{b}+\Lambda_{D}(m_{c};m_{q})>M_{B}.
\label{Dconstraint}   \end{eqnarray} This allowed region is displayed in
Fig. 6 for a light  quark  mass $m_{q}=250$ MeV.  These conditions may
be understood by considering the form of the damping factor in
(\ref{df}). They ensure that the $D^{\ast}$ is below  threshold to
produce a free charm quark and an unphysical particle with mass
$\Lambda_{B}$, and that the $B$ is below threshold to  produce a free
bottom quark and an unphysical particle with mass $\Lambda_{D}$.
(Other conditions such as  $m_{c}+\Lambda_{B}>M_{D}$ and
$m_{b}+\Lambda_{D^{\ast}}>M_{B}$ are less restrictive.)  As we have
said, the existence of an allowed region is expected due to the fact
that the meson masses are fixed.

Our canonical choice $m_{c}=1.44$ GeV and $m_{b}=4.80$ GeV lies   close
to a line running down the middle of the allowed region.  The
corrections to the $B\rightarrow  D^{\ast}$ form factors increase as
the point ($m_{c},m_{b}$) gets closer to the constraint
(\ref{Dstconstraint}) involving $M_{D^{\ast}}$, and the corrections to
the $B\rightarrow  D$ form factors increase as the point gets closer to
the constraint (\ref{Dconstraint}) involving $\Lambda_{D}$.  This effect
manifests itself as the anticorrelation in the full model predictions
for the branching ratios, as seen in Fig. 5. As another illustration we
show  in Fig. 7 the corrections to
$h_{A_{1}}(1)$ and $h_{+}(1)$ as functions of $m_{c}$ with $m_{b}=4.8$
GeV held fixed.  We again see the anticorrelation in the corrections.

It is of interest that for no reasonable choices of quark masses are
the corrections to $h_{A_{1}}(1)$ and $h_{+}(1)$ small simultaneously.
In \cite{HS2} we associated these corrections with hyperfine splitting
effects.  This may be re-phrased in the language of Fig.
6 with the aid of  Table 1 in \cite{HS2}.  There it is
seen that both $M_{D^{\ast}}-\Lambda_{B}$ and $M_{B}-\Lambda_{D}$
decrease  when hyperfine mass splitting is turned off.  From
(\ref{Dstconstraint}) and  (\ref{Dconstraint}) this has the effect of
enlarging the allowed region in the $m_{c}$-$m_{b}$ plane, and the
corrections are correspondingly reduced.

It may seem most reasonable for the corrections to $h_{A_{1}}(1)$
and $h_{+}(1)$ to be of the same order of magnitude, and thus both in
the 10\% to 15\% range.  A more precise statement awaits more
experimental input, as we have described.  The main point of this
paper has been to obtain definite predictions for other quantities,
and how these predictions fare will determine how seriously this model
should be taken.

\vspace{3ex}

\noindent {\bf ACKNOWLEDGMENT} \vspace{1ex}

This research was supported in part by the Natural Sciences and
Engineering Research Council of Canada.  B.H. thanks A. Falk and M.
Luke for useful discussions; he also thanks the Aspen Center for
Physics  where this work was completed.
\vspace{3ex}

\noindent {\bf APPENDIX}
\vspace{1ex}

We summarize in this Appendix the definitions of all relevant
quantities.
The full four-fold decay distribution in the cascade decay
$\overline{B^{o}} \rightarrow D^{\ast +}(\rightarrow D \pi)
+\ell^{-}+\overline{\nu}_{\ell}$ may be written as \cite{KS}
\begin{equation}
\frac{d\Gamma(\overline{B^{o}} \rightarrow D^{\ast +}
(\rightarrow D \pi)\ell^{-}\overline{\nu}_{\ell})}
{d\omega\,d\cos\theta\,d\cos\theta^{\ast}\,d\chi}
= B(D^{\ast +}\rightarrow D\pi)
\sum_{i}\frac{f_{i}(\theta,\theta^{\ast},\chi)}{2\pi}\,
\frac{d\Gamma_{i}(\omega)}{d\omega}.
\end{equation}
$\theta$ is the polar angle of the lepton measured with respect to
the
$D^{\ast}$-direction in the $(\ell \overline{\nu}_{\ell})$ CM system,
$\theta^{\ast}$ is the polar angle of the $D$ relative to the
$D^{\ast}$ in the $D^{\ast}$ rest frame, and
$\chi$ is the azimuthal angle between the two decay planes spanned
by
$(D\pi)$ and $(\ell \overline{\nu}_{\ell})$.  See \cite{KS,ARGUS} for
diagrams.  $B(D^{\ast +}\rightarrow D\pi)$ is the
branching ratio $\Gamma_{D^{\ast +}\rightarrow D\pi}/
\Gamma_{D^{\ast +}\rightarrow {\rm all}}$. The zero lepton mass
approximation has been used.
The angular functions $f_{i}$ are listed in (\ref{table2}).
\begin{equation} \begin{array}{ccc}
i&f_{i}&\hat{H}_{i}  \\ \hline
U&\frac{9}{32}(1+\cos^{2}\theta) \sin^{2}\theta^{\ast}&
|H_{+}|^{2}+|H_{-}|^{2} \\
L&\frac{9}{8}\sin^{2}\theta\,\cos^{2}\theta^{\ast}& |H_{o}|^{2} \\
T&-\frac{9}{16}\sin^{2}\theta\, \sin^{2}\theta^{\ast}
\,\cos 2\chi & {\rm Re}(H_{+}H_{-}^{\ast}) \\
I&-\frac{9}{16}\sin 2\theta\,\sin2\theta^{\ast}\,\cos \chi&
\frac{1}{2}{\rm Re}(H_{+}H_{o}^{\ast}+H_{-}H_{o}^{\ast}) \\
P&\frac{9}{16}\cos\theta\,\sin^{2}\theta^{\ast}&
|H_{+}|^{2}-|H_{-}|^{2} \\
A&-\frac{9}{8}\sin\theta\, \sin 2\theta^{\ast}\,\cos \chi&
\frac{1}{2}{\rm Re}(H_{+}H_{o}^{\ast}-H_{-}H_{o}^{\ast})
\end{array}\label{table2} \end{equation}

The partial helicity rates $d\Gamma_{i}/d\omega$ are given by
\begin{equation}
\frac{d\Gamma_{i}}{d\omega}=\frac{G_{F}^{2}|V_{cb}|^{2}}{48\pi^{3}}
M_{1}M_{2}^{2}\sqrt{\omega^{2}-1}(1+r^{2}-2r\omega)\hat{H}_{i}
(\omega),
\end{equation}
where $M_{1,2}$ are the $\overline{B^{o}}$ and $D^{\ast}$ masses
and
$r=M_{2}/M_{1}$.  $\hat{H}_{i}$ are bilinear expressions
of the three helicity amplitudes $H_{+}$, $H_{-}$ and $H_{o}$
describing the
current-induced transitions $B\rightarrow D^{\ast}$, and are listed in
(\ref{table2}).  A set of four form factors may be defined by
\begin{equation}
\left\langle{D^{\ast}|\;\overline{c}\gamma_{\mu}b\;|
\overline{B^{o}}}\right\rangle =  \sqrt {M_{1}M_{2}}{h}_{V}(
\omega   ){\varepsilon }_{ \mu \upsilon \rho \sigma } {\varepsilon
}_{
2}^{\ast \upsilon }{ v}_{  2}^{\rho }  { v}_{1}^{ \sigma },
\label{formfacs0} \end{equation}
\begin{eqnarray}
\lefteqn{\left\langle{D^{\ast}|\;\overline{c}\gamma_{\mu}\gamma_{
5}b\;|
\overline{B^{o}}}\right\rangle = } \nonumber \\
 && -i\sqrt {M_{1}M_{2}}\left[{( \omega +  1){h}_{{A}_{1}}(
\omega   ){
\varepsilon }_{2 \mu }^{ \ast } -  \left({{h}_{{A}_{2}}( \omega   ){
v}_{1
\mu } +  {h}_{{A}_{3}}( \omega   ){ v}_{2 \mu }}\right) {\varepsilon
}_{
2}^{\ast }\cdot { v}_{  1}}\right],
\label{formfacs1}
\end{eqnarray}
where $\varepsilon_{2}$ is the $D^{\ast}$ polarization vector.
The three helicity amplitudes are given in terms of the four form
factors by
\begin{eqnarray}
H_{\pm}&=&\sqrt{M_{1}M_{2}}(\omega+1)\{h_{A_{1}}\mp [(\omega-
1)/
(\omega+1)]^{1/2}h_{V}\} \nonumber \\
H_{o}&=&\frac{\sqrt{M_{1}M_{2}}(\omega+1)}{\sqrt{1+r^2-
2r\omega}}
\{(\omega-r)h_{A_{1}}-(\omega-1)(h_{A_{3}}+rh_{A_{2}})\}.
\end{eqnarray}

The differential decay rate for $\overline{B^{o}}\rightarrow D^{+}+
\ell+\overline{\nu}_{\ell}$ is given by \cite{KS2}
\begin{equation}
\frac{d\Gamma(\overline{B^{o}}\rightarrow D^{+}
\ell\overline{\nu}_{\ell})}
{d\omega\,d\cos\theta}
=\frac{3}{4}(1-\cos^{2}\theta)\frac{d\Gamma}{d\omega}
\end{equation}
where
\begin{equation}
\frac{d\Gamma}{d\omega}=\frac{G_{F}^{2}|V_{cb}|^{2}}{48\pi^{3}}
M_{1}M_{2}^{2}\sqrt{\omega^{2}-1}(1+r^{2}-2r\omega)
|H_{o}^{D}(\omega)|^{2}.
\end{equation}
$M_{2}$ is now the $D$ mass. The standard pair of form factors are
defined by
\begin{equation}
\left\langle{D|\;\overline{c}\gamma_{\mu}b\;|\overline{B^{o}}}\right
\rangle =
\sqrt{M_{1}M_{2}}\left[h_{+}(\omega)(v_{1}+v_{2})_{\mu}
+  h_{-}(\omega)(v_{1}-v_{2})_{\mu}\right].
\label{formfacs2}
\end{equation}
The amplitude $H_{o}^{D}$ is given in terms of $h_{\pm}$ by
\begin{equation}
H_{o}^{D}=\frac{\sqrt{M_{1}M_{2}}\sqrt{w^{2}-1}}
{\sqrt{1+r^2-2r\omega}} \{(1+r)h_{+}-(1-r)h_{-}\}.
\end{equation}

The total of six form factors have the following values in the heavy
quark limit:
\begin{equation}
h_{V,A_{1},A_{3},+}= \xi \;\;\; ; \;\;\;
h_{A_{2},-}= 0 .
\end{equation}
In all of the model results quoted we include short distance QCD
corrections
$\beta_{i}(\omega)$ \cite{NeuQCD} according to
\begin{equation}
h_{i}(\omega)^{{\rm QCD}}=
h_{i}(\omega)^{{\rm no\,QCD}}+\beta_{i}(\omega)\xi(\omega).
\end{equation}

We implement the momentum cut according
to the prescription set out in \cite{KS}; in our notation, this reads
\begin{equation}
-1\leq \cos\theta \leq {\rm min}(\cos\theta(\omega;p_{\rm cut}),1)
\end{equation}
where
\begin{equation}
\cos\theta(\omega;p_{\rm cut})=\frac{1-r\omega-2p_{\rm
cut}/M_{1}}
{r\sqrt{w^{2}-1}}.
\end{equation}
When converting data into values for branching ratios, the  events lost
due to the cut must be restored. In order to facilitate this we plot in
Fig. 8 the full model prediction for the fraction of
leptons having momentum greater than the cut, for $B\rightarrow D$ and
$B\rightarrow D^{\ast}$.  This fraction is quite insensitive to $m_{c}$.

In uncut form, the forward-backward asymmetry in the angle
$\theta$ is
given by
\begin{equation}
A_{FB}=-\frac{3}{4}\frac{\Gamma_{P}}{\Gamma_{U}+\Gamma_{L}},
\end{equation}
where
\begin{equation}
\Gamma_{i}=\int_{1}^{\omega_{\rm
max}}d\omega\,\frac{d\Gamma_{i}}
{d\omega}\;\;\; ; \;\;\; \omega_{\rm max}=\frac{1+r^{2}}{2r}.
\end{equation}
The lepton momentum cut excludes events in the
extreme backward direction $\cos\theta\rightarrow 1$;  following
\cite{KS} we remove this bias by symmetrizing the definition of
$A_{FB}$ with the forward hemisphere restricted by
\begin{equation}
-{\rm min}(\cos\theta(\omega;p_{\rm cut}),1) \leq \cos\theta \leq 0.
\end{equation}

In uncut form, the $D^{\ast}$ polarization parameter is given by
\begin{equation}
\alpha=2\frac{\Gamma_{L}}{\Gamma_{U}}-1.
\end{equation}

\newpage \noindent{\bf FIGURE CAPTIONS} \vspace{6ex}

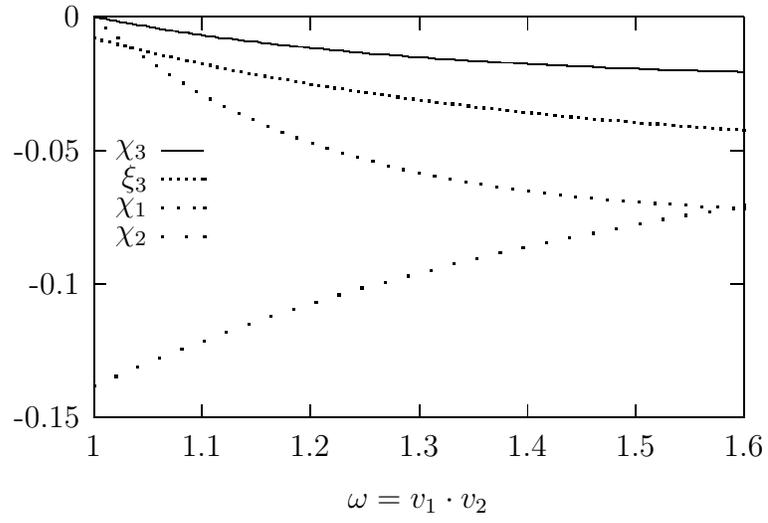
\noindent {\bf FIG. 1:} First-order model predictions for universal
functions
$\xi_{3}$ and $\chi_{1,2,3}$ for $\Lambda=667$ MeV , $m_{q}=250$  MeV,
$g=-0.13$, and $h=0.19$.\vspace{4ex}

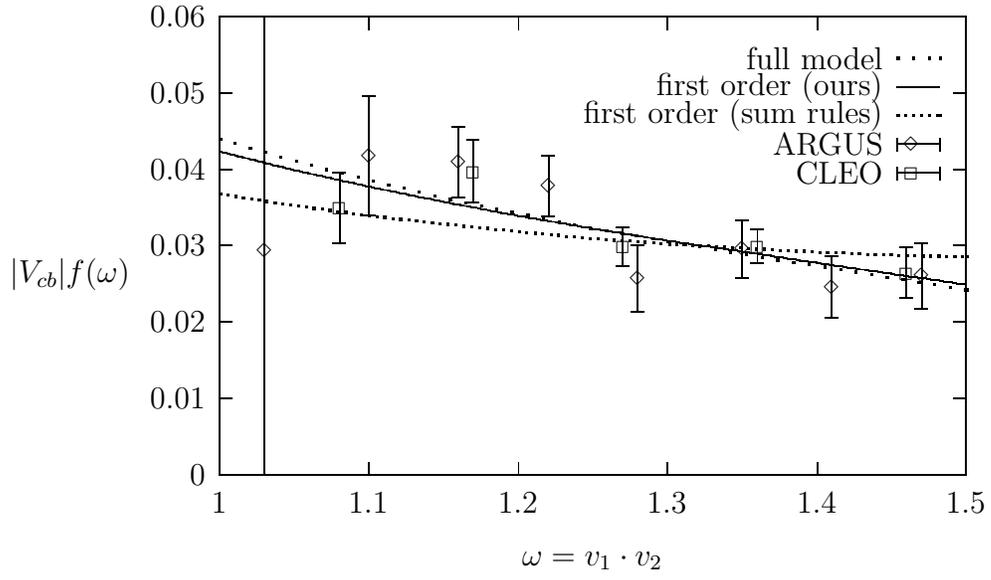
\noindent {\bf FIG. 2:} Predictions for the $B\rightarrow D^{\ast}$
spectrum
$|V_{cb}|f(\omega)$ (equal to $|V_{cb}|\xi(\omega)$ in the heavy  quark
limit).  The respective values of $|V_{cb}|$ are .042 in our first
order model, .037 in first-order QCD sum rules and .038 in our full
model.\vspace{4ex}

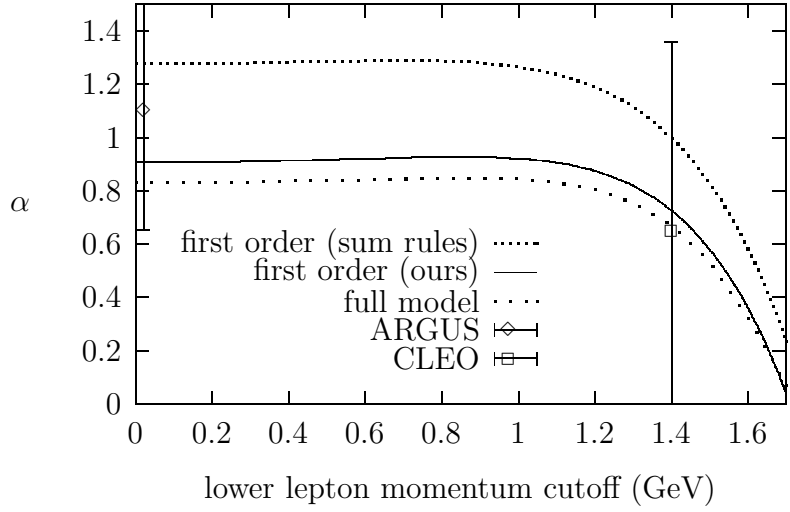
\noindent {\bf FIG. 3:} $D^{\ast}$ polarization parameter $\alpha$ as a
function of  the experimental lower lepton momentum cut $p_{\rm
cut}$.\vspace{4ex}

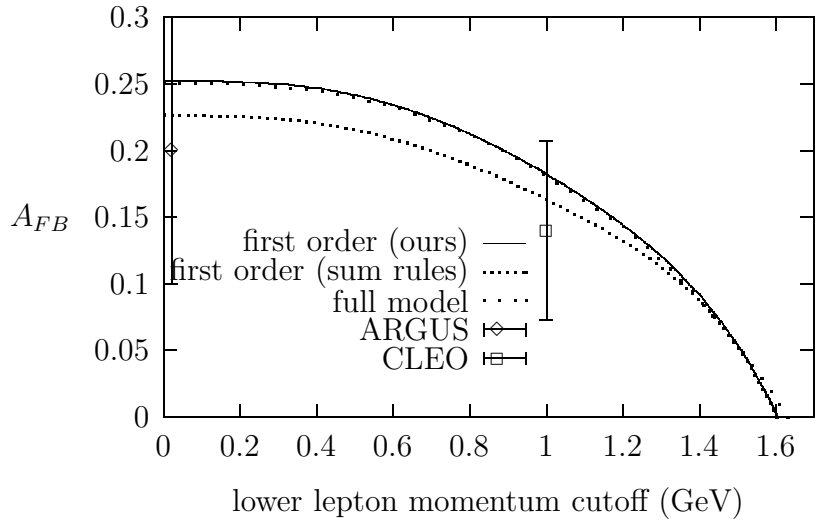
\noindent {\bf FIG. 4:} Forward-backward asymmetry $A_{FB}$ in the
angle
$\theta$, as a function of the lepton momentum cut.  The cut is
performed symmetrically as explained in the text.\vspace{4ex}

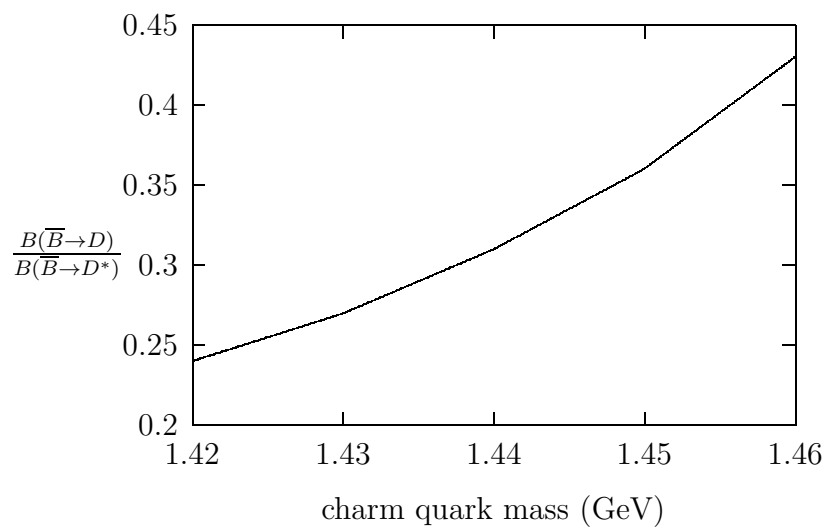
\noindent {\bf FIG. 5:} Full model prediction for
$B(B\rightarrow D)/B(B\rightarrow D^{\ast})$ as a function of the charm
quark mass with bottom quark mass fixed at  4.8 GeV.\vspace{4ex}

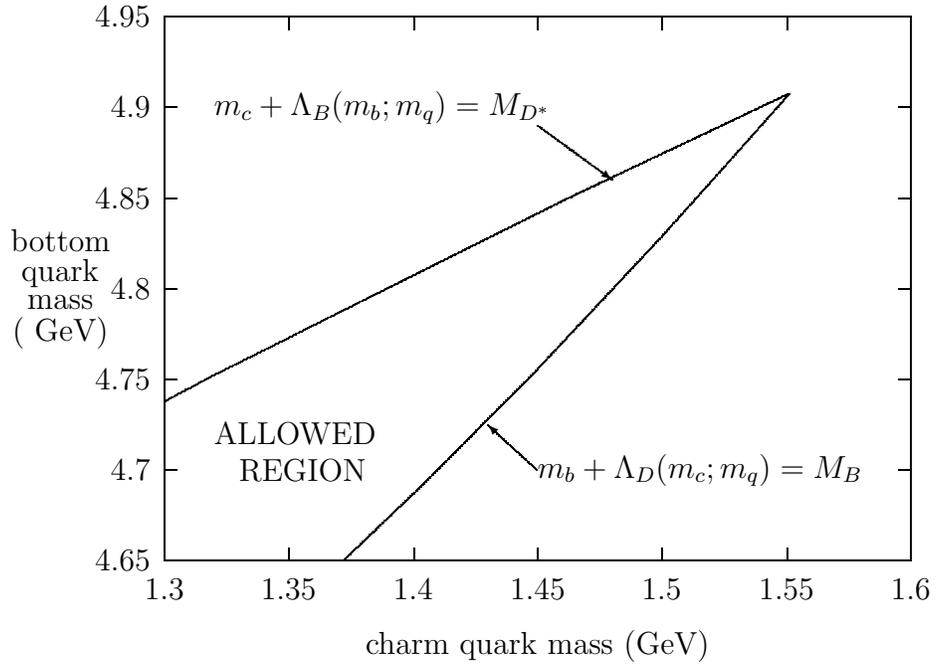
\noindent {\bf FIG. 6:} Allowed region in the $m_{c}$-$m_{b}$ plane,
for light quark mass $m_{q}=250$ MeV.\vspace{4ex}

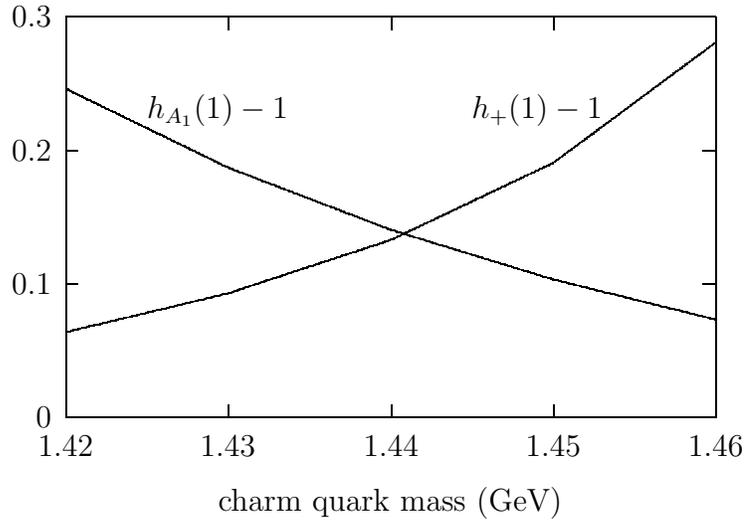
\noindent {\bf FIG. 7:} Full model results for $h_{A_{1}}(1)$ and
$h_{+}(1)$ as functions of the charm quark mass with bottom quark mass
fixed at  4.8 GeV.\vspace{4ex}

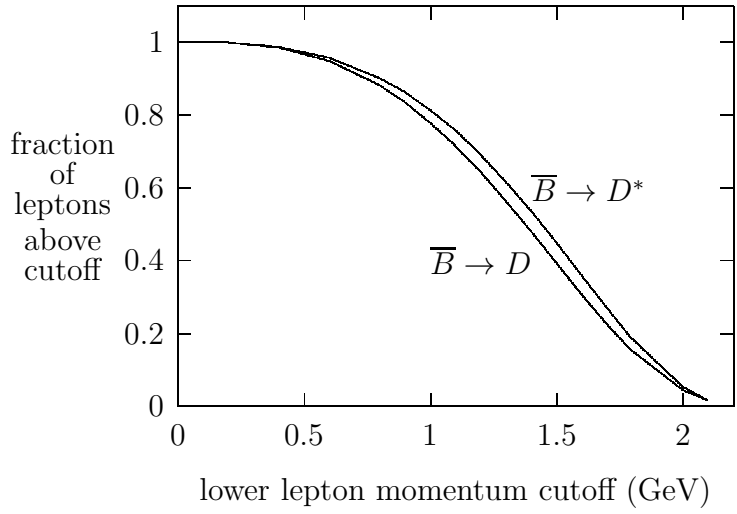
\noindent {\bf FIG. 8:} Full model prediction for fraction of leptons
having momentum greater than a given minimum value.

\begin{thebibliography}{99}
\bibitem{HS1} B. Holdom and M. Sutherland, Phys. Rev. D {\bf 47}, 5067
(1993).
\bibitem{HS2} B. Holdom and M. Sutherland, University of Toronto Report
No. UTPT-92-24, hep-ph/9212277, 1992 (to be published in  Phys. Lett.
B).
\bibitem{Geo} H. Georgi, Phys. Lett. B {\bf 240}, 447 (1990).
\bibitem{IW} N. Isgur and M.B. Wise, Phys. Lett. B {\bf 232}, 113
(1989);  {\it ibid.} B {\bf 237}, 527 (1990).
\bibitem{Luke} M.E. Luke, Phys. Lett. B {\bf 252}, 447 (1990).
\bibitem{NeuRieck} M. Neubert and V. Rieckert, Nucl. Phys. {\bf B382},
97  (1992).
\bibitem{FalkNeu1} A.F. Falk and M. Neubert, Phys. Rev. D {\bf 47},
2965 (1993).
\bibitem{HS5} B. Holdom and M. Sutherland, University of Toronto Report
No. UTPT-93-26, 1992 (unpublished).
\bibitem{FalkNeu2} A.F. Falk and M. Neubert, Phys. Rev. D {\bf 47},
2982 (1993).
\bibitem{Neu1} M. Neubert, Phys. Lett. B {\bf 264}, 455 (1991).
\bibitem{ISGW} N. Isgur, D. Scora, B. Grinstein and M. Wise, Phys. Rev.
D {\bf 39}, 799 (1989).
\bibitem{BSW} M. Wirbel, B. Stech and M. Bauer, Z. Phys. C {\bf 29},
637 (1985).
\bibitem{KS2} J.G. K\"{o}rner and G.A. Schuler, Z. Phys. C {\bf 38},
511 (1988);  {\it ibid.} C {\bf 41}, 690 (1989) (E); {\it ibid.} C {\bf
46}, 93 (1990).
\bibitem{Neurevue} M. Neubert, SLAC Report No. SLAC-PUB-6263, 1993 (to
be published in Physics Reports).
\bibitem{NeuQCD} M. Neubert, Phys. Rev. D {\bf 46}, 3914 (1992).
\bibitem{FNL} A. Falk, M. Neubert and M. Luke, Nucl. Phys. {\bf B388},
363  (1992).
\bibitem{NeuSR} M. Neubert, Phys. Rev. D {\bf 45}, 2451 (1992).
\bibitem{ARGUS} ARGUS Collab., DESY Report No. DESY 92-146, 1992 (to
be published  in the Proceedings of the 26th International Conference
on High  Energy Physics, Dallas, Texas, 1992).
\bibitem{Borto} D. Bortoletto and S. Stone, Phys. Rev. Lett. {\bf 65},
2951 (1990).
\bibitem{CLEO89} CLEO Collab., Phys. Rev. Lett. {\bf 63}, 1667 (1989).
\bibitem{CLEO} CLEO Collab., Phys. Rev. D {\bf 47}, 791 (1993).
\bibitem{KS} J.G. K\"{o}rner and G.A. Schuler, Phys. Lett. B {\bf 226},
185 (1989).
\end{thebibliography}
\end{document}